\documentclass[conference]{IEEEtran}
\IEEEoverridecommandlockouts
\usepackage{cite}
\usepackage{amsmath,amssymb,amsfonts}
\usepackage{algorithmic}
\usepackage{graphicx}
\usepackage{textcomp}
\usepackage{xcolor}
\usepackage{float}
\usepackage{fancyhdr} 
\usepackage{algorithm}
\usepackage{algorithmic}
\usepackage{dblfloatfix}
\makeatletter
\newcommand\fs@norules{\def\@fs@cfont{\bfseries}\let\@fs@capt\floatc@ruled
	\def\@fs@pre{}%
	\def\@fs@post{}%
	\def\@fs@mid{\kern3pt}%
	\let\@fs@iftopcapt\iftrue}
\makeatother
\floatstyle{norules}
\restylefloat{algorithm}\usepackage{algorithm}
\usepackage{algorithmic}

\fancypagestyle{firstpage}{
	\fancyhf{} 
} 
\def\BibTeX{{\rm B\kern-.05em{\sc i\kern-.025em b}\kern-.08em
    T\kern-.1667em\lower.7ex\hbox{E}\kern-.125emX}}
\begin{document}

\title{PDFInspect: A Unified Feature Extraction Framework for Malicious Document Detection\\
}
\author{
	\textbf{S P Sharmila$^{1,2,*}$, and Aruna Tiwari$^{1}$}\\
	\small$^{1}$ Indian Institute of Technology Indore, Madhya Pradesh, India. \\
	$^{2}$Siddaganga Institute of Technology, Tumakuru, Karnataka, India.\\
	E-mail: \{phd2201101012, artiwari\}@iiti.ac.in, 
	sharmila@sit.ac.in  
}

\maketitle
\thispagestyle{firstpage} 
\begin{abstract}
	The increasing prevalence of malicious Portable Document Format (PDF) files necessitates robust and comprehensive feature extraction techniques for effective detection and analysis. This work presents a unified framework that integrates graph-based, structural, and metadata-driven analysis to generate a rich feature representation for each PDF document. The system extracts text from PDF pages and constructs undirected graphs based on pairwise word relationships, enabling the computation of graph-theoretic features such as node count, edge density, and clustering coefficient. Simultaneously, the framework parses embedded metadata to quantify character distributions, entropy patterns, and inconsistencies across fields such as author, title, and producer. Temporal features are derived from creation and modification timestamps to capture behavioral signatures, while structural elements including, object streams, fonts, and embedded images, are quantified to reflect document complexity. Boolean flags for potentially malicious PDF constructs (e.g., JavaScript, launch actions) are also extracted. Together, these features form a high-dimensional vector representation (170 dimensions) that is well-suited for downstream tasks such as malware classification, anomaly detection, and forensic analysis. The proposed approach is scalable, extensible, and designed to support real-world PDF threat intelligence workflows.

\end{abstract}


\section{Introduction}
Portable Document Format (PDF) files have become a ubiquitous medium for digital document exchange due to their platform independence, rich formatting capabilities, and support for embedded multimedia and interactive elements\cite{
 van2009adobe}. However, PDFs are a frequent target for malicious exploitation. 
Attackers often embed JavaScript, execute launch actions, or manipulate metadata to deliver payloads, execute phishing attacks, and exploit vulnerabilities in PDF readers \cite{
 castiglione2010security}, resulting in evasion through AI-based intrusion methods \cite{Sharmila2022}. Consequently, accurate and interpretable analysis of PDF internals is critical for developing robust malware detection systems.
Traditional PDF analysis techniques \cite{ulucenk2011techniques, zhang2018mlpdf} typically focus on content extraction, metadata parsing, and a low-level structural inspection in isolation. While each modality provides valuable insight, relying on a single feature category often fails to capture the multifaceted nature of PDF-based threats. For example, malicious behavior manifest through unusual metadata patterns, obfuscated JavaScript, or even anomalous textual structures. To overcome these limitations, an integrated approach is needed that combines information from multiple layers of a PDF file to generate a comprehensive behavioral signature.

In this work, we present PDFInspect a  unified framework for extracting a rich set of features from PDF documents. Our method combines three core components: (1) Text-to-Graph Conversion, where text content is parsed to construct undirected graphs and compute graph-theoretic metrics; (2) Metadata and Structural Parsing, where key-value fields, timestamps, and document objects are analyzed for suspicious patterns; and (3) Character Composition Analysis, where statistical properties such as entropy, digit counts, and symbol frequencies are quantified across relevant text fields.
The graph-based component captures the topological and lexical relationships between tokens, enabling the detection of obfuscated and sparsely linked text patterns often found in evasive malware. Metadata extraction identifies anomalies in creation dates, font usage, and field consistency. Boolean flags are used to detect the presence of critical components like /JavaScript, /OpenAction, and /URI, which are frequently exploited for code execution.
By fusing these diverse features into a unified vector representation, our system facilitates robust downstream tasks such as classification, clustering, and anomaly detection of PDF files. Furthermore, the approach is designed to be modular and extensible, allowing researchers and practitioners to adapt it for novel threat detection models. 

\section{Background and Related Work}
\label{Background}
PDF documents have long been exploited as vectors for malicious payloads due to their flexibility and complexity. Numerous studies have explored static and dynamic analysis approaches for detecting such threats, with static methods being particularly appealing for their speed and non-intrusive nature. This section reviews the existing key contributions in three relevant areas: PDF malware detection, feature engineering in static analysis, and graph-based approaches for document analysis.
The early works on PDF malware detection relied heavily on rule-based systems and signature matching. 
Tools like PDF Examiner 
and PDFid\cite{StevensPDFTools} extract known suspicious keywords and object references such as /JavaScript, /Launch, and /OpenAction to flag potentially harmful documents. However, these methods are limited by their reliance on predefined patterns, making them susceptible to evasion techniques such as obfuscation \cite{Sharmila2023, Sharmila2025} and object reordering.
To address these limitations, machine learning-based models were introduced.  The use of static features like object counts and metadata fields \cite{aggarwal2006integrating} to train classifiers have demonstrated to be capable of distinguishing malicious from benign files. Moreover, works like Hidost \cite{vsrndic2016hidost} employed hierarchical representations of the PDF structure to learn complex decision boundaries. Despite their improved accuracy, these models often treat features independently and fail to capture deeper structural and relational patterns inherent in text or objects.
Feature engineering\cite{falah2021improving} plays a critical role in PDF malware detection. Commonly used features include document size, number of pages, font usage, object streams, and metadata timestamps. The presence and absence of specific keywords and object types (e.g., /XFA, /URI, /AcroForm) are also powerful indicators of suspicious behavior.
Some studies have proposed statistical descriptors such as entropy and character composition (e.g., counts of digits, uppercase letters, or non-ASCII symbols) to model obfuscation or encoding artifacts. These techniques are particularly effective against adversarial examples\cite{
maiorca2019towards} that use camouflage in text and metadata fields. However, most approaches overlook the latent structural information in document content, such as the relationship between tokens and  sequences of actions encoded in object references.
Graph-based representations have gained popularity in cybersecurity \cite{yan2023graph, sikos2023cybersecurity} and document forensics due to their ability to model relationships and dependencies. Graph models-based PDF analysis is explored \cite{chen2025graph} for representing object references, call graphs, and even document workflows. 
For example,  
modeling PDF internal object references as graphs to detect anomalies in the structure  \cite{weigand2022creating}.
In parallel, NLP tasks\cite{mills2013graph} have utilized text-based graphs 
to extract semantic relationships. These models can be adapted for PDF content by converting extracted text into graphs and computing topological features such as average degree, density, and clustering coefficient. Such features provide insights into the structural regularity and complexity of the document, which can help detection of obfuscated and  machine-generated PDFs.

While previous works have significantly advanced PDF malware detection using static features and structural analysis, most approaches treat features in isolation and ignore the potential synergy between content, metadata, and graph topology. Our ``PDFInspect" bridges the gap by combining character composition, metadata flags, and graph-theoretic features derived from document text. This holistic approach captures both local and global patterns in the PDF file, enhancing the system's robustness against evasive and polymorphic threats.
\section{Proposed Methodology}
\label{sec:Proposed}
\begin{figure}[h]
	\centering
	\includegraphics[width=9cm,height=6cm, scale=0.4]{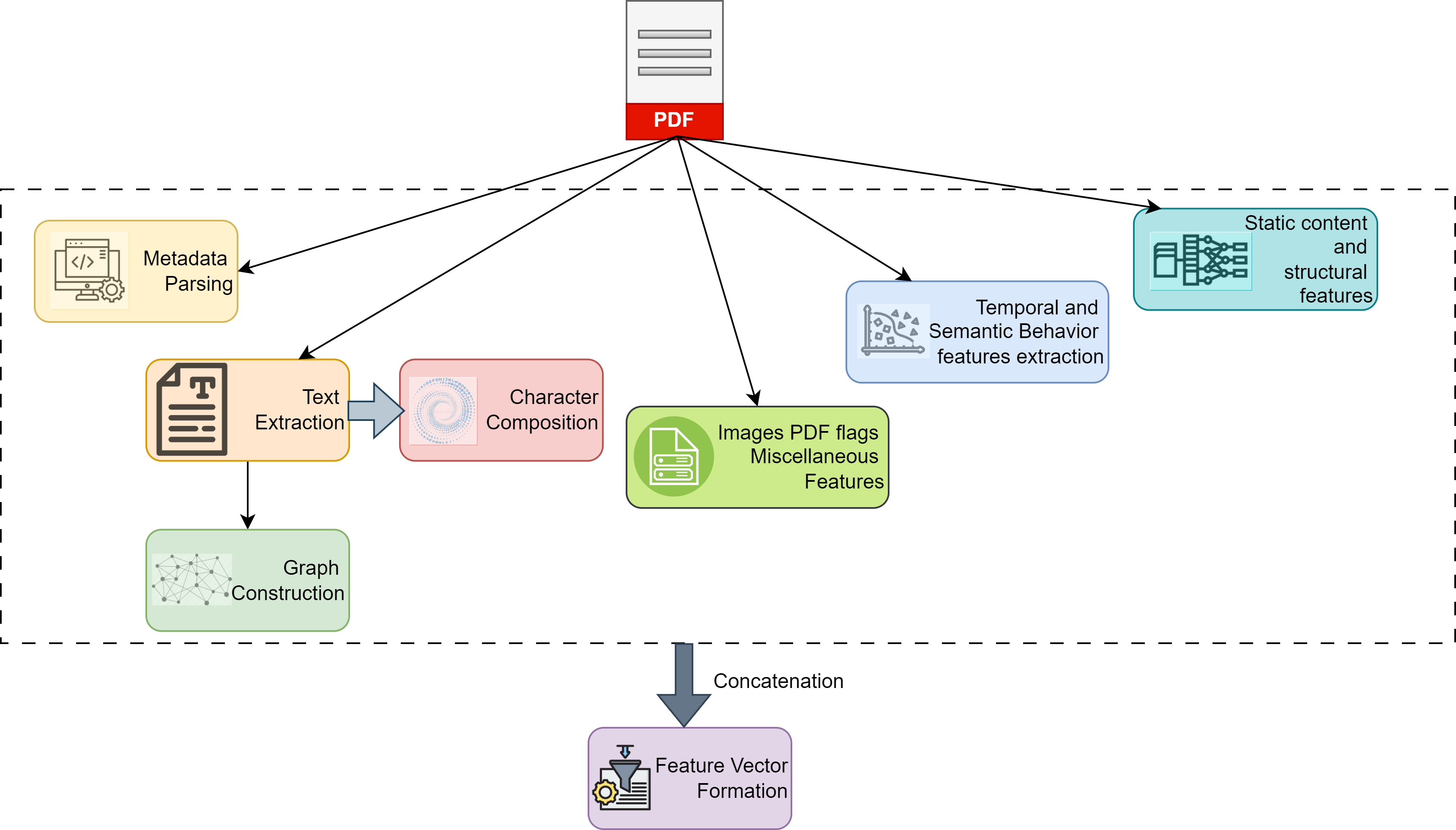}
	\caption{Complete Pipeline of proposed PDFInspect Feature Extraction module }
	\label{Fig:PipelineFeatureExtraction}
\end{figure}

The PDFInspect system integrates multiple analytical perspectives to extract a high-dimensional feature vector for each PDF file. These features encompass content structure, metadata semantics, character-level statistics, and graph-theoretic properties derived from the document's textual content.  
Proposed PDFInspect is a unified system that analyzes PDF files by extracting a diverse set of features across multiple layers. It parses the textual content into graph structures to derive graph-theoretic features, evaluates metadata fields for character composition, entropy, and timestamp irregularities, and quantifies structural components such as the number of objects, streams, fonts, and embedded images. In addition to these, it generates a high-dimensional feature vector for each PDF, capturing both semantic and structural characteristics relevant to malware detection.
The complete pipeline is depicted in Fig. \ref{Fig:PipelineFeatureExtraction}, the framework consists of the following major components, detailed in subsequent sections:\\
1. Textual Content Extraction \& Graph Construction\\
2. Graph-Theoretic Feature Computation\\
3. Metadata and Structural Feature Extraction\\
4. Character Composition and Statistical Analysis\\
5. Temporal and Semantic Behavior features extraction\\
6. Static content and structural features

\begin{table*}
	\caption{Details of dataset Collection (with proposed feature set) in comparison with existing datasets}
	\label{tab:Dataset}
	\centering
	\begin{tabular}{|c|c|c|c|c|c|c|l|}
				\hline
		Sl.&Existing &Features and feature count	 & \# Malicious & \# Benign & Total \\ \hline
		1.&CIC\cite{issakhani2022pdf} &General +structural (32)&21291&	9109&28791\\ \hline
		
		2.&Contagio\cite{chenette2009malicious}&Structural(135)&10814&7396&18210\\ \hline
		
		3.&PDFRep\cite{liu2023evaluating}&Graph                      (8)&22761&10477&33238\\ \hline
		
		4.&Feature engg\cite{yerima2022malicious}&Feature engineered metadata(41)&21291&9109&28791\\ \hline
		
		5.&Proposed&Metadata + graph + structural                        (170)&128876 &133237&262113 \\ \hline
		\end{tabular}
\end{table*}
\subsection{Textual Content Extraction and Graph Construction}

PDF files often embed both visible and hidden text. In this phase we perform Text Extraction, Tokenization, Graph Formation.
In \textbf{Text Extraction}, all textual content is extracted from each page of the PDF.  
In \textbf{Tokenization}, the extracted text is splitted into words(tokens), which are preprocessed (e.g., lowercased, stripped of punctuation) to standardize representation. Finally, in 
\textbf{Graph Formation}, an undirected word graph is constructed for each page; where, Nodes represent unique tokens, Edges represent co-occurrence within a fixed-size window(e.g., 2 or 3 words). 
Multiple pages result in multiple graphs aggregating their features. 
For \(d\) number of numeric features, \(m\) number of string-based features and \(n\) number of boolean indicator features, let a PDF file $F$ be processed into a feature vector $\mathbf{x} \in \mathbb{R}^d \cup \text{Str}^m \cup \{0, 1\}^n$, where,\\
\noindent
-$\mathbb{R}^d$ contains numeric features (e.g., counts, ratios, time deltas)

\noindent
-$\text{Str}^m$ contains metadata strings (lowercase, raw)

\noindent
-$\{0, 1\}^n$ contains boolean presence flags (e.g., timezone present)

\noindent
Then, Text Parsing for Graph representation is modeled as follows:

\noindent
For text $T = \bigcup_{i=1}^{N} T_i$, where $T_i$ is text from page $i$,
we define a graph:
$
G = (V, E) \text{ where, }\\ E = \{(u, v) \mid \text{a line in } T \text{ contains exactly 2 words: } u, v\}
$

\subsection{ Graph-Theoretic Feature Computation}
The structural complexity of a document’s text is captured by computing graph-theoretic metrics from page-wise word graphs, where each page is represented as a network of unique words(nodes) and their connections(edges). For every page, we extract and aggregate the node count ($|V|$, the number of unique words), edge count ($|E|$, the number of token connections), and graph density ($D$). Additionally, document-level features are constructed by calculating the mean and maximum values of three central graph properties: average degree, which reflects the mean number of edges per node; clustering coefficient, which measures the tendency of a node’s neighbors to be interconnected; and degree centrality, indicating the influence of a node based on its connection count. Together, these features reveal patterns of regularity, repetition, and topological structure in the text, offering signals of automation, obfuscation, or code injection. The resulting graph feature vector $\mathbf{g}$ for each document consists of five core metrics;  
formally expressed as $\mathbf{g} = \left[|V|, |E|, \bar{d}, D, C \right] \in \mathbb{R}^5$, providing a compact representation of the document’s structural properties. Average degree, Density and Clustering coefficient are represented by Eq. (\ref{eq.AvgDeg}), (\ref{eq.Density}) and (\ref{eq.ClusCo}) respectively.\\
\begin{equation}
	\small
	\label{eq.AvgDeg}
	\bar{d} = \frac{1}{|V|} \sum_{v \in V} \deg(v)
\end{equation}
\begin{equation}
	\small
	\label{eq.Density}
	D = \frac{2|E|}{|V|(|V| - 1)}	
\end{equation}
\begin{equation}
	\small
	\label{eq.ClusCo}
	C = \frac{1}{|V|} \sum_{v \in V} \frac{\# \text{triangles through } v}{\# \text{connected triples centered at } v}
\end{equation}
\subsection{Metadata and Structural Feature Extraction}
This module targets embedded metadata and document-level structural elements that are often manipulated in malicious PDFs.
Key metadata fields such as `/Author', `/Title', `/Creator', `/Producer', `/Subject', `/Keywords', `/CreationDate', and `/ModDate' are extracted. 
Boolean Flags for malicious indicators and
presence of suspicious entries are flagged by `/JavaScript', `/OpenAction', `/Launch', `/URI', `/RichMedia', `/AcroForm', `/EmbeddedFile', `/XFA', etc.
Structural counts include
number of objects, fonts, pages, image streams, embedded files,
version of the PDF (e.g., 1.4, 1.7),
Object stream and XRef usage patterns.
Timestamp consistency can be
extracted and compared by `/CreationDate' and `/ModDate'.
We compute Temporal Delta 
$
\Delta_t = |\text{ModDate} - \text{CreationDate}|
$.
This captures inconsistencies, manipulations, and behavioral artifacts in document structure.
To extract Metadata feature vector, let, metadata fields be $M = \{m_1, ..., m_k\} \subset \{\text{author, title, ...}\}$.
For each string $s_i = m_i$, we extract $\mathbf{f}_i^{\text{xxx}}$ as in Eq.(\ref{eq.Fi}). Each field contributes to frame $\mathbf{f}_i$ as in Eq.(\ref{eq:FiAll}) to form entire metadata fields composed by Eq.(\ref{eq:Fmeta}).
\begin{equation}
	\small
	\label{eq.Fi}
	\begin{aligned}
		f_i^{\text{dot}} &= \text{count of `.' in } s_i \\
		f_i^{\text{len}} &= |s_i| \\
		f_i^{\text{num}} &= \sum \mathbf{1}(c \in 0\text{-}9) \\
		f_i^{\text{oth}} &= \sum \mathbf{1}(c \notin a\text{-}zA\text{-}Z0\text{-}9) \\
		f_i^{\text{uc}} &= \sum \mathbf{1}(c \in A\text{-}Z)
	\end{aligned}
\end{equation}
\begin{equation}
	\small
	\label{eq:FiAll}
	\mathbf{f}_i = \left[f_i^{\text{dot}}, f_i^{\text{len}}, f_i^{\text{num}}, f_i^{\text{oth}}, f_i^{\text{uc}}\right] \in \mathbb{R}^5
\end{equation}
\begin{equation}
	\small
	\label{eq:Fmeta}
	\mathbf{f}_{\text{meta}} = \bigcup_{i=1}^k \mathbf{f}_i \in \mathbb{R}^{5k}
\end{equation}

\subsection{Character Composition and Statistical Analysis}
To capture obfuscation and encoding tricks, character-level analysis is applied to relevant fields (`/Title', `/Author', etc.,) and the main text body.
For Entropy calculation,
Shannon entropy is computed to assess randomness with $p_i$, the probability of character $i$ in the string as in Eq.(\ref{eq:entropy}):
\begin{equation}
	\small
	\label{eq:entropy}
	H = -\sum p_i \log_2(p_i)
\end{equation}
Character-level counts include,
number of digits, uppercase letters, lowercase letters, whitespace characters.
number of special characters and non-ASCII characters. For field length and word count, text summary metrics considered are,
average word length, maximum token length, token diversity (ratio of unique to total tokens).
These features contribute to detect encoded payloads, randomized strings, and fields that mimic benign structures.

\subsection{Temporal and Semantic Behavior Features}
These features relate to sequence, time-order, behavioral dependencies in the content; typically, modeled via graph and sequential methods, reflecting execution-like and text-flow dynamics.
Temporal features include
page-wise object sequencing,
text token co-occurrence graphs, 
structural flow (e.g., sequence of action triggers),
graph-theoretic measures (degree, centrality, entropy of graph).
Let,
$T_c$ be $\text{creation timestamp}$, 
$T_m$ be $\text{modification timestamp}$,
we define temporal feature $
\delta_t$ as in Eq.(\ref{eq:deltaT}) and Boolean flags $b_c$ and $b_m$ as in Eq.(\ref{eq:boolean}). So, finally temporal features are as in Eq.(\ref{eq:Temporal}). 
\begin{equation}
	\small
	\label{eq:deltaT}
	\delta_t = 
	\begin{cases}
		(T_m - T_c).total\_seconds() & \text{if both exist} \\
		0 & \text{otherwise}
	\end{cases}
\end{equation}
\begin{equation}
	\small
	\label{eq:boolean}
	b_c = \mathbf{1}(``Z" \in T_c), \quad b_m = \mathbf{1}(``Z" \in T_m)
\end{equation}
\begin{equation}
	\small
	\label{eq:Temporal}
	\mathbf{f}_{\text{time}} = \left[\delta_t, b_c, b_m\right] \in \mathbb{R} \times \{0,1\}^2
\end{equation}

\subsection{Static Content and Structural Features}
These are features are extracted without execution of the file; relying on PDF’s internal structure, metadata, and embedded objects.
It includes structural counts like,
number of objects, pages, streams, fonts and positions of /JavaScript, /OpenAction, etc., object tree depth and
Byte offsets of suspicious elements.
File-level features comprises of file size,
compression ratio, object entropy,
number of embedded files or scripts, and
encoding or encryption used.
PDF keyword flags comprises of 
presence of /JavaScript, /OpenAction, /AA, /Launch, usage of /Encrypt, /RichMedia, /URI and frequency of potentially malicious keywords.
Image-based features includes number of images, embedded image formats, image size, resolution, steganographic potential indicators.
From raw page content $P_i$ and binary data, for counting
`endobj', `stream', `endstream', `/Font', `images', etc.,
position stats (min, max, avg) are modeled by Eq.(\ref{eq.PosStat}). So, for each feature-type $t$, we define Eq.(\ref{eq.Ft}). Summing all structural features for $s$ is the total structural dimension as in Eq.(\ref{Fstruct}).
\begin{equation}
\small \label{eq.PosStat}
\text{pos}_{\text{obj}} = 
\left\{ \min_i p_i,\; \max_i p_i,\; \operatorname*{avg}_{\substack{i}} p_i \right\}
\end{equation}
\begin{equation} 
	\small
	\label{eq.Ft}
	\mathbf{f}_t = \left[\text{count}_t, \text{min pos}_t, \text{max pos}_t, \text{avg pos}_t\right] \in \mathbb{R}^4
\end{equation}
\begin{equation} 
	\small
	\label{Fstruct}
	\mathbf{f}_{\text{struct}} = \bigcup_t \mathbf{f}_t \in \mathbb{R}^{s}
\end{equation}
Similarly, for Image-based features,
let $I_i$ be number of images on page $i$, the following equations are used to classify scanned vs synthesized documents.
\begin{equation}
	\small
	\text{total\_images} = \sum_i |I_i| 
\end{equation}
\begin{equation}
	\small
	\text{image\_mismatch} = \mathbf{1}(\text{image count} \neq \text{image pixel area logic})
\end{equation}
For File-level features,
`file\_size' $\in \mathbb{N}$,
`PDF version' can be converted to float if needed, and `count\_page' $\in \mathbb{N}$
forming simple scalar values.
For PDF keyword flags, from the trailer/root dictionary, flags are set 
for suspicious tags like `/JS', `/Launch', `/URI', etc.
\begin{equation}
	\small
	f_k = \begin{cases}
		1 & \text{if key } k \text{ is found in Root} \\
		0 & \text{otherwise}
	\end{cases}
\end{equation}
So the final  boolean vector framed as:
\begin{equation}
	\small
	\mathbf{f}_{\text{pdf\_flags}} \in \{0, 1\}^m
\end{equation}
	
\subsection{Image Feature Extraction}
Let a PDF contain $M$ embedded images indexed by $j=1,\dots,M$. For each image $j$, with:
$w_j, h_j \in \mathbb{N}$ be its pixel width and height,
$a_j = w_j \cdot h_j$ be its pixel area,
$o_j \in [0, \text{size}(F)]$ be a byte offset of the image object within the file.
Choosing pixel-area thresholds $\tau_{\text{xs}} < \tau_{\text{s}} < \tau_{\text{m}} < \tau_{\text{l}}$ (implementation constants) to bucket image sizes, then we extract the following counts as in Eq.(\ref{eq:imageCount}):
\begin{equation}
		\small
		\label{eq:imageCount}
		\begin{aligned}
			\text{count\_image\_total} &= M\\
			\text{count\_image\_xsmall} &= \sum_{j=1}^{M} \mathbf{1}\!\left(a_j < \tau_{\text{xs}}\right)\\
			\text{count\_image\_small}  &= \sum_{j=1}^{M} \mathbf{1}\!\left(\tau_{\text{xs}} \le a_j < \tau_{\text{s}}\right)\\
			\text{count\_image\_med}    &= \sum_{j=1}^{M} \mathbf{1}\!\left(\tau_{\text{s}} \le a_j < \tau_{\text{m}}\right)\\
			\text{count\_image\_large}  &= \sum_{j=1}^{M} \mathbf{1}\!\left(\tau_{\text{m}} \le a_j < \tau_{\text{l}}\right)\\
			\text{count\_image\_xlarge} &= \sum_{j=1}^{M} \mathbf{1}\!\left(a_j \ge \tau_{\text{l}}\right)
		\end{aligned}
\end{equation}
Pixel-area aggregates and normalization ratios are computed as in Eq.(\ref{eq:PixAggre}) and Eq.(\ref{eq:NorRatio}). Further, byte-offset statistics of image positions are obtained by Eq.(\ref{eq:OffPos}).  
\begin{equation}
	\small
		\label{eq:PixAggre}
		\text{image\_totalpx} \;=\; \sum_{j=1}^{M} a_j \;=\; \sum_{j=1}^{M} (w_j h_j)
\end{equation}
\begin{equation}
	\small
		\label{eq:NorRatio}
		\text{ratio\_imagepx\_size} \;=\; \frac{\text{image\_totalpx}}{\text{size}(F)}
\end{equation}
\begin{equation}
		\small
		\label{eq:OffPos}
		\begin{aligned}
			\text{pos\_image\_min} &= \min_{1\le j\le M} o_j\\
			\text{pos\_image\_max} &= \max_{1\le j\le M} o_j\\
			\text{pos\_image\_avg} &= \frac{1}{M}\sum_{j=1}^{M} o_j \quad (\text{if } M>0;\; \text{else } 0).
		\end{aligned}
\end{equation}
Putting these in the exact order commonly used in the headers along with a sanity flag \(image\_mismatch\) for consistency check as in Eq.(\ref{eq:F_image}), we get a 12-dimensional block. 
\begin{equation}
		\small
		\label{eq:F_image}
		\small
		\mathbf{f}_{\text{image}} =
		\big[
		\text{count\_image\_xsmall},\;
		\dots
		\text{pos\_image\_max},\;
		\text{image\_mismatch}
		\big]
\end{equation}
Finally, \(\mathbf{f}_{\text{misc}}\) is composed of heterogeneous auxiliary features to catch all feature set for file-level, entropy, encoding, and anomaly indicators that don’t belong strictly to meta/time/struct/image/pdf-flag groups, but are critical for robustness.
\subsection{Feature Vector Formation}
All extracted features from the above modules are concatenated into a final fixed-length feature vector per PDF sample.
The final vector (over 170) per file  as in Eq.(\ref{eq:vectorFile}) such that $\mathbf{x}_F \in \mathbb{R}^{d_1} \cup \text{Str}^{d_2} \cup \{0,1\}^{d_3}$:
\begin{equation}
	\small
		\label{eq:vectorFile}
		\mathbf{x}_F = \left[
		\mathbf{g}, \mathbf{f}_{\text{meta}}, \mathbf{f}_{\text{time}}, \mathbf{f}_{\text{struct}}, \mathbf{f}_{\text{image}}, \mathbf{f}_{\text{pdf\_flags}}, \mathbf{f}_{\text{misc}}
		\right]
\end{equation}
This vector $\mathbf{x}_F$ becomes a row in dataset matrix $X \in \mathbb{R}^{N \times D}$
This feature vector is designed to feed downstream machine learning or deep learning models for:
Binary classification (Malicious vs Benign),
Multi-class classification (Malware family detection),
Anomaly detection (for zero-day or evasive samples). Finally, this combined feature extractor turns each PDF into a numerical signature, that blends	Graph theory, Text statistics, Temporal analysis, and File structure parsing; making it suitable for machine learning pipelines, especially for PDF malware detection.
\section{ Experimental Details with Results and discussion}
To evaluate the effectiveness of the proposed feature engineering framework for PDF malware detection, we evaluated this dataset on significant state-of-the-art models. In this section, we describe the   
evaluation of our dataset using state-of-the-art learning models, with baseline comparison.
\subsection{ Dataset Description}
The dataset used in this study consists of 262,113 PDFs, carefully curated to include a balanced representation of both malicious (128,876) and benign (133,237) samples. The malicious files were aggregated from diverse repositories such as VirusTotal, VirusShare, Malware Bazaar, URLHaus, and the Canadian Institute for Cybersecurity (CIC), ensuring broad coverage of real-world threat variants.  Benign samples, on the other hand, were sourced from trustworthy domains including government publications, academic papers, e-books, and open-source document archives, thereby reflecting typical non-malicious document usage. The composition of our dataset is as shown in Table \ref{tab:Dataset}. To ensure dataset integrity, duplicate files were removed based on hash comparisons, and all samples were verified through multiple antivirus engines to confirm ground-truth labeling.  
To mitigate the inherent distribution shift that can emerge from noisy, irregular, or obfuscated documents commonly found in malware datasets, we have applied two distribution-normalization strategies:
(1) Content-agnostic feature equalization, where structural and behavioral features were standardized to reduce reliance on superficial formatting cues, and
(2) Augmented benign sampling, which injected additional benign files from heterogeneous, real-world user-generated sources and synthetic perturbations to better match the variability seen in malicious PDFs.
These steps ensured that the classifier learned threat-specific signatures rather than stylistic or formatting artifacts.
\begin{table*}[h]
	\caption{Classification Performance (Proposed Feature Set)}
	\label{tab:Performance}
	\centering
	\begin{tabular}{|c|c|c|c|c|c|l|}
		
		\hline
		Model &Accuracy(\%) & Precision & Recall & F1-Score & AUC-ROC & CK Score\\ \hline
		
		Random Forest    & 97.5    & 0.973    & 0.978  & 0.975    & 0.986&96.2   \\ \hline
		XGBoost          & 98.1    & 0.979     & 0.984  & 0.981    & 0.990 &95.8  \\ \hline
		SVM (RBF kernel) & 96.3    & 0.961     & 0.965  & 0.963    & 0.972 &95.6  \\ \hline
		Feedforward ANN  & 97.8    & 0.976     & 0.980  & 0.978    & 0.988 & 95.4 \\ \hline
		\textbf{KAN} &\textbf{99.30} &\textbf{99.52} &\textbf{99.56 }&\textbf{99.30} &\textbf{0.993} &\textbf{98.5} \\ \hline
		
	\end{tabular}
\end{table*}
\begin{figure*}[ht]
	\centering
	\begin{minipage}{0.2\textwidth}
		\centering
		\includegraphics[width=4cm, height=3.8cm]{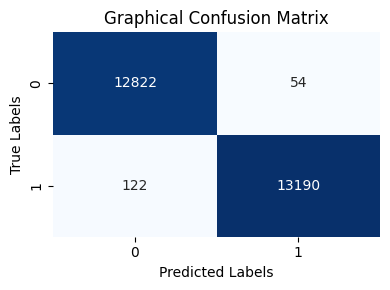}
	\end{minipage}
	\hspace{0.5cm}
	\begin{minipage}{0.2\textwidth}
		\centering
		\includegraphics[width=4cm, height=3.8cm]{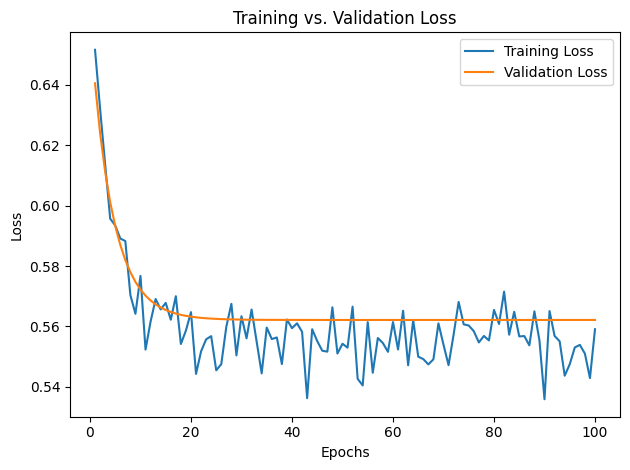}
	\end{minipage}
	\hspace{0.5cm}
	\begin{minipage}{0.2\textwidth}
		\centering
		\includegraphics[width=4cm, height=3.8cm]{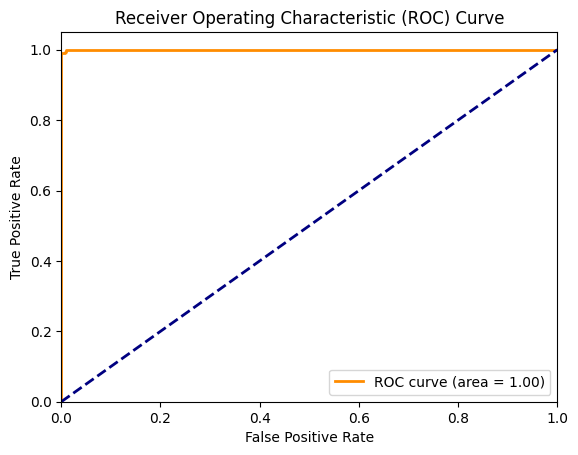}
	\end{minipage}
	\hspace{0.5cm}
	\begin{minipage}{0.2\textwidth}
		\centering
		\includegraphics[width=4cm, height=3.8cm]{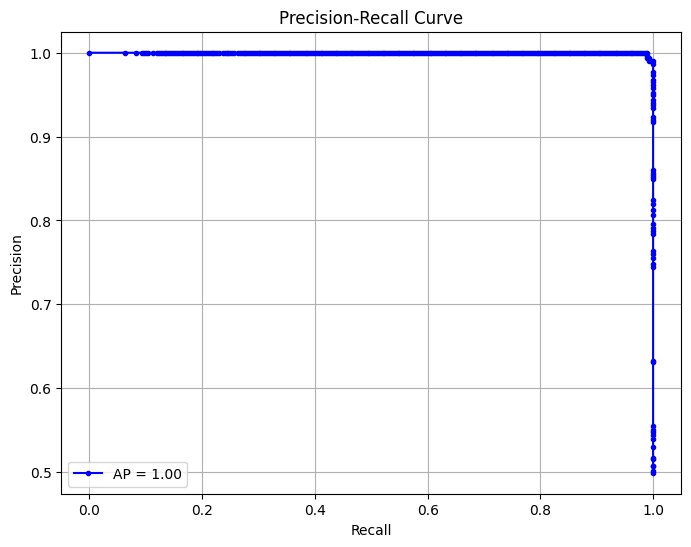}
	\end{minipage}
	\caption{Confusion Matrix, Training vs validation losses, ROC-AUC and PR-Curve of the Best performing Model}
	\label{fig:BestResults}
\end{figure*}
\subsection{  Learning Models}
To evaluate the discriminative power of the extracted features, we evaluated our dataset with supervised classifiers: Random Forest,
Gradient Boosted Trees(XGBoost),
Support Vector Machines(SVM),
Feedforward Neural Network(MLP) and Kolmogorov-Arnold Networks(KAN).
KAN\cite{KAN} is a novel neural network which is emerging as a competitor to MLP. All models are trained on 80:10:10 ratio with cross-fold validation. 
Our KAN consists of 4-Layers and 5-Fold\cite{SharmilaKAN2025} cross-validation. The experiments were conducted using Python 3.10, leveraging a suite of powerful libraries for document processing, feature extraction, and machine learning. Key libraries include PyMuPDF and PyPDF2 for parsing and handling PDF files, NetworkX for graph-based analysis, and scikit-learn and XGBoost for model training and evaluation. Data manipulation and numerical computation were efficiently managed using pandas and NumPy. All computations were performed on an Intel Core i7 ($11^{\text{th}}$ Gen) system with 32 GB RAM, running on Ubuntu 22.04 LTS. The feature extraction process was highly optimized, with an average execution time of approximately 0.1 seconds per PDF document. The entire detection pipeline, including classification, was able to process each PDF in under one second, demonstrating both the efficiency and scalability of the proposed method.
Accuracy, Precision, Recall, F1-Score, AUC-ROC, PR Curve, and Cohen's Kappa Score(CK Score) are the metrics considered for comparison.
\begin{table}[H]
	\caption{Comparison with Baseline Models (Accuracy \%)}
	\label{tab:compBase}
	\begin{tabular}{|c|c|c|l|}
		\hline
		Model           & Baseline 1 & Baseline 2 & PDFInspect \\ \hline 
		Random Forest   & 96.1       & 91.2       & 97.5           \\ \hline
		XGBoost         & 90.3       & 92.8       & 98.1            \\ \hline
		SVM(RBF Kernel)&95.3 & 97.1 &97.9\\ \hline
		Feedforward ANN & 95.5     & 91.9       & 97.8            \\ \hline
		\textbf{KAN} &\textbf{96.4}& \textbf{96.6}& \textbf{99.3}\\ \hline
	\end{tabular}
\end{table}
\subsection{Baseline Comparison}
To validate the utility of our feature set, we compared it with two baseline feature sets:\\
\noindent
Baseline 1: Structural features only (counts of objects, pages, and flags)\\
Baseline 2: Metadata + JavaScript flags without text or graph features

Our method consistently outperformed both baselines, especially in identifying obfuscated and evasive malicious PDFs.
All experiments were conducted using deterministic random seeds and version-controlled environments. 
The classification performance of each model is summarized in Table \ref{tab:Performance}, 
demonstrating that models trained on the proposed feature set achieve superior performance across all evaluation metrics compared to those trained on baseline features.
Among all models, KAN performed the best with an accuracy of 99.3\%, an F1-score of 99.3\%, and an AUC of 0.993 and XGBoost being the second-best performer, showcasing its ability to handle the nonlinear relationships among the diverse set of features.

We compared our combined feature-based approach against the two baseline methods detailed in Table \ref{tab:compBase}. 
The proposed approach improves accuracy by 3-4\% over baseline methods. This significant boost confirms that incorporating graph-based, text structure and character entropy features, captures meaningful patterns that are otherwise ignored in traditional approaches. Confusion Matrix, Training vs validation losses, ROC-AUC and PR-Curve of the best performing model KAN are as shown in Fig. \ref{fig:BestResults}.  This diverse set of influential features underscores the multi-perspective nature of the proposed method, where both structural anomalies and content complexity are indicative of malicious intent.

\subsection{ Advantages of the Proposed Method}
PDFInspect offers notable advantages to enhance effectiveness of documents analysis. \textbf{Multi-view analysis} is one of the key strengths that improves detection accuracy by capturing structural, semantic, and behavioral perspectives. The proposed method demonstrates \textbf{resilience to evasion techniques}, leveraging graph-based features and entropy metrics to identify obfuscations and sophisticated evasion attempts. The solution is \textbf{parallelizable}, allowing efficient analysis over large datasets without compromising performance. 
While MLPdf\cite{zhang2018mlpdf} overfits to content entropy. However, a metadata+graph+structural trio model resists via multi-view invariance: metadata rules block blatant evasion, graphs features preserve topology under perturbations, and structural features  provide fallback counts; totally yielding 10-20\% better robustness than baselines. Overall, these advantages make PDFInspect a robust, flexible, and well-suited tool for modern document analysis and threat detection tasks. 
\section{Conclusion}
\label{Conclusion}
This work demonstrates a unified feature engineering approach PDFInspect, which combines textual, structural, graph-theoretic, and statistical perspectives, significantly enhancing the detection of malicious PDFs. We proposed a PDF feature extraction module and our dataset comprises of over 170 features extracted on more than 100 thousand samples using the proposed method. Our dataset is evaluated by state-of-the-art AI models, which found that the proposed dataset is more favourable for the application of deep neural networks. The proposed method is also found to be scalable and adaptable, making it suitable for integration into enterprise-level email gateways, antivirus engines, and digital forensic pipelines particularly those most favorable with deep learning models. The combination of graph-theoretic and entropy-based features significantly enhanced detection capability. Future directions include integrating dynamic observation, such as monitoring document interactions, analyzing execution flows, and tracking contextual changes over time, while considering system aspects and scalability concerns. 
Working in this direction enhances the system’s capacity to detect subtle threats, reducing misclassifications and guiding the refinement of detection strategies against evolving document and malware tactics.


\begin{thebibliography}{00}
\tiny

\bibitem{van2009adobe} Van der Knijff, Johan. ``Adobe portable document format." \textit{Inventory of long-term preservation risks}, v0 2, pp 20-56. (2009)
\bibitem{castiglione2010security} Castiglione, Aniello, Alfredo De Santis, and Claudio Soriente. ``Security and privacy issues in the Portable Document Format." \textit{Journal of Systems and Software} 83, no. 10, pp 1813-1822. (2010)
\bibitem{Sharmila2022}
Sharmila, S. P., Pratyush Shukla, and Narendra S. Chaudhari. ``A distinguished method for network intrusion detection using random initialized viterbi algorithm in hidden Markov model." In 2022 \textit{OITS International Conference on Information Technology (OCIT)}, pp. 273-277. IEEE, 2022.

\bibitem{ulucenk2011techniques} Ulucenk, Caglar, Vijay Varadharajan, Venkat Balakrishnan, and Udaya Tupakula. ``Techniques for analysing pdf malware." In \textit{18th Asia-Pacific Software Engineering Conference}, pp. 41-48. IEEE, 2011.
\bibitem{zhang2018mlpdf} Zhang, Jason. ``MLPdf: an effective machine learning based approach for PDF malware detection." \textit{arXiv preprint} arXiv:1808.06991 (2018).	
\bibitem{StevensPDFTools} Didier Stevens. 2011. PDF Tools. https://blog.didierstevens.com/programs/pdftools/. Last accessed: 11 Aug., 2025.

\bibitem{Sharmila2023}
Sharmila, S. P., Aruna Tiwari, and Narendra S. Chaudhari. ``Obfuscated malware detection using multi-class classification." In \textit{IEEE International Conference on Cloud Computing in Emerging Markets (CCEM)}, pp. 170-175. IEEE, 2023.

\bibitem{Sharmila2025}
Sharmila, S. P., Shubham Gupta, Aruna Tiwari, and Narendra S. Chaudhari. ``Leveraging memory forensic features for explainable obfuscated malware detection with isolated family distinction paradigm." \textit{Computers and Electrical Engineering} 123 (2025): 110107.



\bibitem{aggarwal2006integrating}
Aggarwal, Ashish, and Pankaj Jalote. ``Integrating static and dynamic analysis for detecting vulnerabilities." In 30th \textit{Annual International Computer Software and Applications Conference (COMPSAC'06)}, vol. 1, pp. 343-350. IEEE, 2006. 
\bibitem{vsrndic2016hidost} Šrndić, Nedim, and Pavel Laskov. ``Hidost: a static machine-learning-based detector of malicious files." \textit{EURASIP Journal on Information Security}, no. 1 : 22. (2016)

\bibitem{chen2025graph} Chen, Yufan, Ruiping Liu, Junwei Zheng, Di Wen, Kunyu Peng, Jiaming Zhang, and Rainer Stiefelhagen. ``Graph-based Document Structure Analysis." \textit{arXiv preprint} arXiv:2502.02501 (2025).
\bibitem{chenette2009malicious} 
Chenette, Stephan. ``Malicious documents archive for signature testing and research-contagio malware dump". (2009).
\bibitem{falah2021improving} Falah, Ahmed, Lei Pan, Shamsul Huda, Shiva Raj Pokhrel, and Adnan Anwar. ``Improving malicious PDF classifier with feature engineering: A data-driven approach." \textit{Future Generation Computer Systems} 115, pp 314-326. (2021)
\bibitem{maiorca2019towards} Maiorca, Davide, Battista Biggio, and Giorgio Giacinto. ``Towards adversarial malware detection: Lessons learned from PDF-based attacks." \textit{ACM Computing Surveys (CSUR)} 52, no. 4, pp  1-36. (2019)
\bibitem{issakhani2022pdf} Issakhani, Maryam, Princy Victor, Ali Tekeoglu, and Arash Habibi Lashkari. ``Pdf malware detection based on stacking learning." In \textit{ICISSP}, pp. 562-570. (2022).
\bibitem{liu2023evaluating} Liu, Ran, Robert Joyce, Cynthia Matuszek, and Charles Nicholas. ``Evaluating Representativeness in PDF Malware Datasets: A Comparative Study and a New Dataset." In 2023 \textit{IEEE International Conference on Big Data (BigData)}, pp. 3017-3024. IEEE, 2023.


\bibitem{mills2013graph} Mills, Michael T., and Nikolaos G. Bourbakis. ``Graph-based methods for natural language processing and understanding—a survey and analysis." \textit{IEEE Transactions on Systems, Man, and Cybernetics}: Systems 44, no. 1, pp 59-71, (2013)

\bibitem{yan2023graph} Yan, Bo, Cheng Yang, Chuan Shi, Yong Fang, Qi Li, Yanfang Ye, and Junping Du. ``Graph mining for cybersecurity: A survey." \textit{ACM Transactions on Knowledge Discovery from Data} 18, no. 2 : pp 1-52. (2023)
\bibitem{sikos2023cybersecurity} Sikos, Leslie F. ``Cybersecurity knowledge graphs." \textit{Knowledge and Information Systems} 65, no. 9, pp 3511-3531. (2023)
\bibitem{weigand2022creating} Weigand, Maximilian, and Alexander Fay. ``Creating virtual knowledge graphs from software-internal data." In \textit{48th Annual Conference of the IEEE Industrial Electronics Society}, pp. 1-6. IEEE, 2022.

\bibitem{yerima2022malicious} Yerima, Suleiman Y., Abul Bashar, and Ghazanfar Latif. ``Malicious PDF detection based on machine learning with enhanced feature set." In \textit{14th International Conference on Computational Intelligence and Communication Networks (CICN)}, pp. 486-491. IEEE, 2022.

\bibitem{KAN} Liu, Ziming, Yixuan Wang, Sachin Vaidya, Fabian Ruehle, James Halverson, Marin Soljačić, Thomas Y. Hou, and Max Tegmark. ``Kan: Kolmogorov-arnold networks." Accepted by \textit{International Conference on Learning Representations (ICLR) 2025}, \textit{arXiv preprint} arXiv:2404.19756 (2024).

\bibitem{SharmilaKAN2025} 
Sharmila, S. P., Shubham Gupta, Aruna Tiwari, and Narendra S. Chaudhari. "Unveiling evasive portable documents with explainable Kolmogorov Arnold Networks resilient to generative adversarial attacks." Applied Soft Computing (2025): 113537.



\end{thebibliography}
\end{document}